\documentclass[12pt]{article}

\usepackage[margin=1in]{geometry}

\usepackage{setspace}           
\usepackage{graphicx}           
\usepackage{amsmath}            
\usepackage{amsfonts}
\usepackage{marginnote}         
\usepackage{datetime}           
\usdate                         
\usepackage{enumitem}           
\usepackage{subfigure}          
\usepackage{rotating}           
\usepackage{hyperref}           
\usepackage{float}              
\usepackage[longnamesfirst]{natbib} 
\usepackage{xcolor}
\usepackage{texintro}           
\usepackage{bibentry}           
\usepackage{longtable}
\usepackage{pdflscape}
\usepackage{everypage}

\newcommand{\Var}{\ensuremath{\mathrm{Var}}}
\newcommand{\Cov}{\ensuremath{\mathrm{Cov}}}
\newcommand{\e}{\epsilon}

\newtheorem{thm}{Theorem}[section]

\newtheorem{prop}[thm]{Proposition}
\newtheorem{defn}[thm]{Definition}

\newcommand{\Lpagenumber}{\ifdim\textwidth=\linewidth\else\bgroup
  \dimendef\margin=0 
  \ifodd\value{page}\margin=\oddsidemargin
  \else\margin=\evensidemargin
  \fi
  \raisebox{\dimexpr -\topmargin-\headheight-\headsep-0.5\linewidth}[0pt][0pt]{%
    \rlap{\hspace{\dimexpr \margin+\textheight+\footskip}%
    \llap{\rotatebox{90}{\thepage}}}}%
\egroup\fi}
\AddEverypageHook{\Lpagenumber}%

\begin{document}

\nobibliography*                                
\setlist{noitemsep}                             


\doublespacing

\newcommand{\abs}{
We propose a novel approach to infer investors' risk preferences from their portfolio choices, and then use the implied risk preferences to measure the efficiency of investment portfolios. We analyze a dataset spanning a period of six years, consisting of end of month stock trading records, along with investors' demographic information and self-assessed financial knowledge.
Unlike estimates of risk aversion based on the share of risky assets, our statistical analysis suggests that the implied risk aversion coefficient  of an investor increases with her wealth and financial literacy. 
Portfolio diversification,  Sharpe ratio, and expected portfolio returns correlate positively with the efficiency of the portfolio, whereas a higher standard deviation reduces the efficiency of the portfolio. We find that affluent and financially educated investors as well as those holding retirement related accounts hold more efficient portfolios.}

\title{Risk Preferences and Efficiency of Household Portfolios}

\author{
  Agostino Capponi
    \thanks{Department of Industrial Engineering and Operations Research, Columbia University, New York, USA, \ 10027, e-mail: ac3827@columbia.edu.}
\and
Zhaoyu Zhang 
    \thanks{Department of Mathematics, University of Southern California, Los Angeles, California,  USA, 90089,  \  e-mail: zzhang51@usc.edu.}
}

\maketitle
\thispagestyle{empty}

\begin{abstract}
  \abs
\end{abstract}







\section{Introduction}\label{sec:intro}

Attitude towards risk is a crucial factor in the investors' decision making process.
In this work, we first propose a novel approach to infer investors' risk preferences from portfolio choices. We then leverage a unique data set of household portfolio positions and corresponding demographics, spanning a six-year time horizon, to shed light on the main determinants of risk preferences and the efficiency of portfolio allocations.

The concept of  portfolio efficiency dates back to the pioneering work of  \cite{Markowitz:1952} on mean-variance analysis. In his work, Markowitz assumes that investors are risk averse, and trade off risk and return of their portfolios. Investors act rationally and, for a given level of risk, they choose the portfolio which yields the highest expected return. Plotting these portfolios in the risk and expected return space traces the efficient frontier (or ``the Markowitz bullet'') in the absence of a risk-free asset. If a risk-free asset is included, the line which is tangent to the efficient frontier at the portfolio with the highest Sharpe ratio 
is called the capital market line (CML). 

According to theory, any rational investor would choose a portfolio on the capital market line depending on her risk preferences. 
 However, using our data set we observe that the majority of investors' portfolios lie under the capital market line (see Figure \ref{fig:CML} for an illustration). This provides supportive evidence that the vast majority of investors do not outperform, and rather underperform, the market. Existing literature has identified several reasons why this may happen: excessive trading (e.g. \cite{Odean:1998} and \cite{Kumar:2009}), overconfidence (e.g. \cite{Barber_Odean:2001}),  and underdiversification (e.g. \cite{Kelly:1995} and  \cite{Von_Gaudecker:2015}).

A large body of literature has investigated the process of inferring risk preferences from data in various environments. \cite{Sahm:2012} used panel data on hypothetical gambles over lifetime income to quantify changes in risk tolerance over time.  \cite{Chiappori_Salanie:2019} identified the distribution of risk preferences among bettors in parimutuel horse races; \cite{Kimball_Sahm:2008} developed a quantitative proxy for risk tolerance based on hypothetical income survey. \cite{Schubert:1964} investigated the effect of gender on risk attitudes in financial decision-making via a designed experiment. \cite{Bucciol_Miniaci:2011} estimated risk preferences from household survey portfolio data, which distinguish between risk free assets, bonds, stocks, human capital, and real estate. {Unlike the above mentioned studies, we focus on the stock market, and consider two types of data sets. The first are investor specific data, which  consist of portfolio positions and demographics of households \citep{Barber_Odean:2000}\footnote{We are grateful to Prof. Terrence Odean for agreeing to share the data set with us.}. This is a rich dataset, which includes end of month trading records, for a period of six years ranging from 1991 through 1996. The second dataset consists of market data used to construct expected values and standard deviations of asset returns based on Fama French factors and industry portfolios.}

We next describe our approach for inferring the risk preferences of an investor. In our work, we project the expected return and standard deviation of an investor's portfolio onto the capital market line. We define the investor's risk aversion coefficient to be that associated with the projection of the portfolio on the capital market line. Such a risk aversion coefficient $\theta$ can be computed explicitly as 
$$
	\theta = \frac{(1 + \lambda_{mkt}^2)/ \sigma_{obs}}{\lambda_{obs} + 1/\lambda_{mkt}},
 $$
and is determined by the market Sharpe ratio ($\lambda_{mkt}$), the observed portfolio Sharpe ratio ($\lambda_{obs}$), and the observed portfolio standard deviation ($\sigma_{obs}$). Consistent with intuition, our formula implies that investors become less risk averse as (i) the portfolio standard deviation increases, and (ii) the portfolio Sharpe ratio increases.

Our proposed approach differs from those used in existing literature, which either use the proportion of risky assets over total wealth (see \cite{Riley:1992}), or instead utilize the portfolio standard deviation (see \cite{Bucciol_Miniaci:2011}) to imply the risk aversion coefficients.

We use the formula developed above to assess the relationship between risk attitudes and individual characteristics, both in the cross section and over time. 
Our findings are consistent with results from earlier literature, and additionally enrich them. In line with existing studies (see  \cite{Bucciol_Miniaci:2011}, \cite{Morin_Suarez:1983}, \cite{Palsson:1996}),  and \cite{Morin_Suarez:1983}), we find that  investors become more risk averse as they get older. 
We also find that investors' risk preferences are sensitive to market characteristics. We use the CBOE Volatility Index (VIX) as a measure of aggregate market risk, and find that investors become more risk averse as the VIX rises.  \cite{Guiso:2018} also find that individual investors become more risk averse following the 2008 crisis, but their results are based on repeated surveys of portfolio choices. 
We also discover a substantial degree of  heterogeneity across investors. Specifically, we find that those who are financially more literate and possessing a higher net worth are more risk averse. These findings stand it contrast with existing literature, where different risk-aversion measures and  datasets have been used. It is worth mentioning here the study of \cite{Riley:1992}, who estimated the relative risk aversion index ratio of risky assets over wealth, \cite{Guiso:2008} who measured risk aversion based on the maximum price a consumer is willing to pay for a risky security, and \cite{Rooij:2011} who measured the risk aversion as the willingness to participate in the stock market. 
We interpret our results by the fact that wealthier and financially more literate investors are better at diversifying risk in the stock market. As a result their portfolio becomes less volatile, and thus their implied risk-aversion is higher. Furthermore, we find that the implied risk aversion of investors holding margin and cash accounts is lower than that of investors holding retirement related accounts only. {This can be understood by the fact that retirement account are ``safer'', in that they yield smaller return and  have lower volatility, hence the implied risk-aversion of these investors is higher.} 

We then turn to investigating the efficiency of investors' portfolios. Unlike  \cite{Gibbons:1989}, who proposed a test of mean-variance efficiency for a given investor's portfolio based
 on the difference between the Sharpe ratios of the actual and the efficient portfolios, 
we measure efficiency by the Euclidean distance between the observed investor's portfolio and its projection on the capital market line. The latter is the ``closest'' to the observed portfolio, among all portfolios selected by an investor with identical risk preferences and lying on the efficient frontier. Hence, the efficiency of the portfolios is not uniquely identified by its Sharpe ratio. If two portfolios have the same Sharpe ratio, the less volatile portfolio would be more efficient according to our measure. We find that older, wealthier, and financially more literate investors tend to hold more efficient portfolios. This result is related to \cite{Calvet:2009} and \cite{Campbell:2006}. {Rather than analyzing portfolio efficiency, these papers study how the wealthiness and financial education of households affects their probability of making financial mistakes.} 
{Our regression analysis also reveals that more diversification (measured by the number of stocks in the invested portfolio), larger expected returns, and lower standard deviation improves the efficiency of the portfolio.} 
{This result can be understood in terms of the fact that any rational investor prefers to maximize return for a given level of risk, or equivalently, minimize risk for a target level of return.}

The paper proceed as follows. In Section \ref{sec:method}, we describe the technique used to imply risk aversion from an observed portfolio, and then define the proposed measure of portfolio efficiency. We describe the data set in Section \ref{sec:data}. We report the details of our implementation procedure in  Section \ref{sec:implement}. Section~\ref{sec:hist} presents the descriptive statistics of our data. Section \ref{sec:regression} reports the findings from our panel data regressions. We conclude in Section~\ref{sec:conclusions}.

\section{Methodology}\label{sec:method}

In this section, we present the methodology used to infer risk preferences from investor portfolio choices, and define the measure of portfolio efficiency used.

\subsection{Implied Risk Preferences}
\begin{figure}[htbp]
\begin{center}
 \includegraphics[width = 0.7\textwidth, height = 7cm]{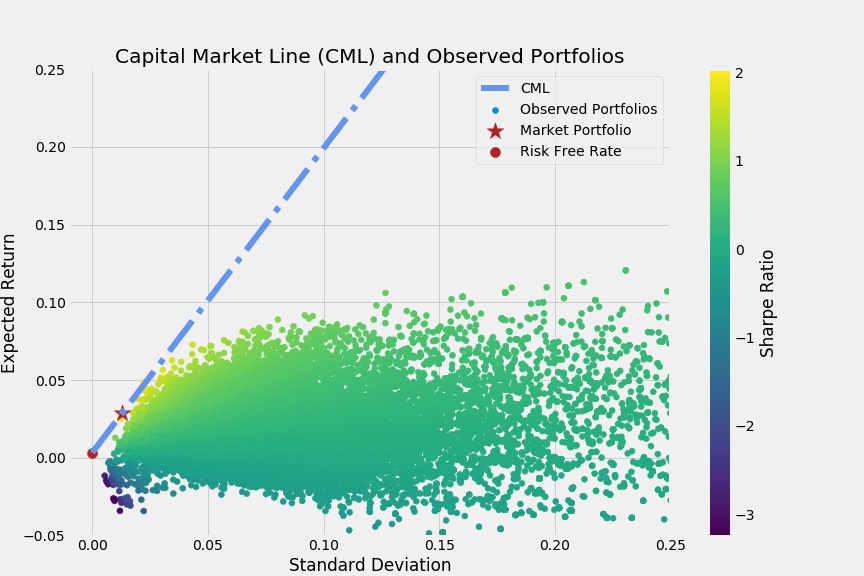}
  \caption{Capital market line and observed investor portfolios using the data set for the month of January 1992. The data used to construct the market portfolio comes from the Fama-French 49-industry portfolio. The expected return and standard deviation for investors' portfolios and the  Fama-French 49-industry portfolios are estimated using a Fama-French five-factor model. The expected return and standard deviation of the market portfolio can be computed explicitly as shown in Section \ref{sec:implement}. We take the risk-free rate from the Fama-French five-factor dataset.}
 \label{fig:CML}
    \end{center}
\end{figure}

Our formulas for implying investors' risk aversion and measuring portfolio efficiency builds on the foundational work of \cite{Markowitz:1952}. According to Markowitz's portfolio criterion, for a given level of risk investors choose the portfolio which yields the highest expected return. Such optimal portfolios, when plotted in the standard deviation and expected return space,  lie on the so-called efficient frontier.
\cite{Tobin:1958} extended Markowtiz's work by adding a risk-free asset. His ``separation theorem" leads a new efficient frontier, called the capital market line. This line is a graphical representation of all efficient portfolios obtained from a linear combination of the risk-free asset, with risk-free rate $r_f$, and the market portfolio, with return $\mu_{mkt}$ and standard deviation $\sigma_{mkt}$. The return $\mu_p$ and standard deviation $\sigma_p$ of any portfolio on the capital market line satisfies:
\begin{equation}
	\mu_p := r_f + \frac{\mu_{mkt} - r_f}{\sigma_{mkt}} \sigma_p =  r_f + \lambda_{mkt} \sigma_p,
\end{equation}
where $$\lambda_{mkt} := \frac{\mu_{mkt} - r_f}{\sigma_{mkt}}$$ is called the Sharpe ratio of the market. 
In Figure \ref{fig:CML}, we plot the capital market line and observed investors' portfolios for the month of January 1992. The data set used to construct the market portfolio comes from the Fama-French 49-industry portfolio.
Each rational investor   chooses a portfolio along the capital market line based on her risk preference. Such a portfolio  maximizes the investor's returns for a given level of risk.  
The existence of a one-to-one correspondence between the risk aversion coefficient and the portfolio on the capital market line can be seen from the mean-variance criterion.  Suppose an investor has risk aversion coefficient $\theta$. Then, he would invest a proportion $w^*$ of her wealth in the market portfolio and the remaining proportion of $(1-w^*)$ in the risk-free asset. The weight $w^*$ which maximizes the mean-variance utility is given by:
$$w^* = \arg \max_w \{ w \mu_{mkt} + (1- w) r_f - \frac{\theta}{2} w^2 \sigma_{mkt}^2\}.$$
From the first order condition, we obtain 
\begin{equation}\label{eqn:w}
    w^* = \frac{\mu_{mkt} - r_f}{\sigma_{mkt}^2 \theta} = \frac{\lambda_{mkt}}{\theta \sigma_{mkt}}.
\end{equation} 
Denote by $(\mu_{obs},\sigma_{obs})$ the expected return and standard deviation of observed portfolio for an investor. Because of the one-to-one correspondence between risk preferences and portfolio choices in \eqref{eqn:w}, we can use the capital market line to infer the investor's risk aversion. This is accomplished by projecting the investor's portfolio onto the capital market line. We then define the implied risk aversion coefficient of the portfolio to be that of the projected portfolio $( \mu_{obs}^{\perp},\sigma_{obs}^{\perp}$). A straightforward calculation shows that the expected return and the standard deviation of the projected portfolio is given by
$$(\mu_{obs}^{\perp}, \sigma_{obs}^\perp) 
= \left(\frac{\lambda_{mkt}^2 \mu_{obs} + \lambda_{mkt} \sigma_{obs} + r_f}{1 + \lambda_{mkt}^2}, \frac{\lambda_{mkt} \mu_{obs} + \sigma_{obs} - \lambda_{mkt} r_f}{1 + \lambda_{mkt}^2} \right).$$
%
The risk aversion coefficient associated with the portfolio with $(\mu_p^{\perp},  \sigma_p^\perp)$ is obtained by matching the volatility of the portfolio with risk aversion $\theta$, given by $w^* \sigma_{mkt} $, with the volatility of the projected portfolio,  $\frac{\lambda_{mkt} \mu_{obs} + \sigma_{obs} - \lambda_{mkt} r_f}{1 + \lambda_{mkt}^2}$. Formally, 
$$w^* \sigma_{mkt} = \frac{\lambda_{mkt} \mu_{obs} + \sigma_{obs} - \lambda_{mkt} r_f}{1 + \lambda_{mkt}^2},$$
and it then follows that the implied risk aversion for an observed portfolio with expected return and standard deviation $(\mu_{obs},\sigma_{obs})$ is given explicitly by 
\begin{equation}\label{eq:rv}
	\theta = \frac{(1 + \lambda_{mkt}^2)/ \sigma_{obs}}{\lambda_{obs} + 1/\lambda_{mkt}}. 
\end{equation} 

\begin{figure}[htbp]
\begin{center}
 \includegraphics[width = 0.6\textwidth, height = 0.6\textwidth]{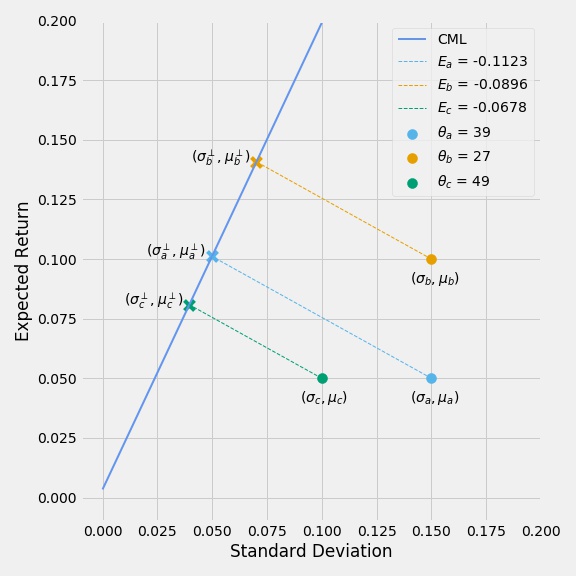}
 \caption{Comparison of investors' $(a,b,c)$ implied risk aversion  coefficients {$\theta$} and portfolio efficiency {$E$} based on the expected return and standard deviation of investors' portfolios.}
 \label{fig:method}
    \end{center}
\vspace{-0.5cm}    
\end{figure}

It is clear from our formula that the investor's risk aversion coefficient depends both on the portfolio mean and the portfolio standard deviation. Moreover, the formula implies that, given two portfolios with the same expected return, the investor who holds the one with the higher standard deviation is less risk averse. This is illustrated in Figure \ref{fig:method}, where investor $a$ with risk aversion coefficient equal to 39 holds a portfolio with expected return and standard deviation of $(0.05, 0.15)$, and investor $c$ with risk aversion coefficient of $27$ holds a portfolio with expected return and standard deviation of $(0.05, 0.1)$. Moreover, it is clear from our formula that the investor's implied risk aversion coefficient would decrease as the mean return of the portfolio increases while the volatility stays the same. As shown in Figure \ref{fig:method}, while maintaining the same standard deviation of 0.15, increasing the expected return from 0.05 to 0.1 (shifting from point $(\sigma_a, \mu_a)$ to $(\sigma_b, \mu_b)$) would yield an increase of the risk aversion coefficient by 10.

\subsection{Measuring the Inefficiency of Portfolios}

In our data, the vast majority of investors' portfolios are not sufficiently close to the capital market line, i.e., they are inefficient. Existing research has provided various explanations for the origins of these inefficiencies. For example, \cite{Barber_Odean:2000} find that households' overconfidence leads to  excessive trading, and households' portfolios are largely under-diversified. As a result, the average household net annual return is much lower than the market returns. 

We next define the measure of efficiency that we use to evaluate household portfolios. 
\begin{defn}
A portfolio is {\it efficient} if its Sharpe ratio is greater than the market's Sharpe ratio. If, instead, the reverse inequality holds, we say that the portfolio is inefficient.
\end{defn}
We measure the efficiency of the portfolio by the Euclidean distance between the observed portfolio and the projected portfolio on the capital market line. 

Our definition of inefficiency is driven by the following considerations. If the portfolio is inefficient, the projected portfolio is the closest ``optimal portfolio'' (in the sense of lying on the efficient frontier) a risk-averse investor can attain by eliminating the excess volatility and increasing the expected return, while maintaining the same risk preference. For example, in Figure \ref{fig:method} an investor with risk aversion coefficient equal to 39 can achieve a ``better'' portfolio, if she moves from $(\sigma_a, \mu_a)$ to $(\sigma_a^{\perp}, \mu_a^{\perp})$. The new portfolio has a standard deviation which is lower by 0.1, and an expected return which is higher by 0.05.  Mathematically, we denote the portfolio efficiency by $E$, and quantify it as

\begin{equation}
E  = 
    \begin{cases}
    \sqrt{(\mu_{obs} - \mu_{obs}^{\perp})^2 + (\sigma_{obs} - \sigma_{obs}^{\perp })^2} & \mbox{  if } \frac{\mu_{obs} - r_f}{\sigma_{obs}} > \frac{\mu_{obs}^{\perp} - r_f}{\sigma_{obs}^{\perp}},\\
       - \sqrt{(\mu_{obs} - \mu_{obs}^{\perp})^2 + (\sigma_{obs} - \sigma_{obs}^{\perp })^2} &  \mbox{  if } \frac{\mu_{obs} - r_f}{\sigma_{obs}} < \frac{\mu_{obs}^{\perp} - r_f}{\sigma_{obs}^{\perp}}.\\
    \end{cases}
\end{equation}

\section{Data}\label{sec:data}
We use three data sets in our analysis: investor portfolios and demographics, Fama French industry portfolios, and Fama French factors.
\subsection{Household Portfolios and Demographics Dataset}
Our data set of investors' portfolio positions and corresponding investor demographics come from \cite{Barber_Odean:2000}. This dataset is provided by a large discount brokerage firm, and it includes the end of month trading records, from January 1991 through December 1996, for the accounts opened by the households as well as the corresponding demographic information, account type, and self-assessed suitability. We filter out  households for which there is missing information (such as missing net worth and income information), or such that  their portfolios consist of assets whose prices are not available in the Center for Research in Security Prices (CRSP) database. We are then left with a dataset of 37,108 households including 54,141 accounts.\footnote{Our original   dataset of positions consists of 128829 accounts. The demographic dataset includes 55432 households. The suitability dataset, which contains self-reported net worth, income, and financial knowledge information, includes 77995 households. The base dataset (containing account information) has 158006 accounts opened by 77995 households.}

Investors' account, demographics,  and suitability information are all recorded at the time investors opened their accounts. They include:
\begin{enumerate}[label=(\arabic*)]
    \item \emph{Net worth proxy}: The net worth of each household is assigned to one of the following six categories based on the dollar amount:  1--24,999, 25,000--49,999, 50,000--74,999, 75,000--99,999, 100,000--249,999, and $>$250,000. The number of households in each of these categories is, respectively, $23817$, $2158$, $1237$, 1442, 4752, and 3702.
    \item \emph{Income proxy}: The income proxy is recorded in one of five categories, depending on the dollar amount of the households' annual income: 1--24,999, 25,000--49,999, 50,000--74999, 75,000--99,999, $>$100,000. We observe, respectively, 23395, 3945, 2798, 3085, and 3885 households for these categories.
    \item \emph{Knowledge}: This variable records self assessed financial knowledge. 3103 accounts have extensive knowledge, 11188 accounts have good knowledge, 7373 accounts have limited knowledge,  2087 accounts have zero knowledge, and the remaining accounts did not report their knowledge information.
    \item \emph{Age}: The age variable records the oldest person's age in a household. It belongs to one of seven groups: 18--24, 25--34, 35--44, 45--54, 55--64, 65--74, $>$ 75. These groups consist of 52, 1843, 8822, 10289, 6746, 5521, and 3835 households respectively.
    \item \emph{Number of children}: A integer from 0 to 6. 26154 households have no child, 6236 households have 1 child, 3525 households have 2 children, 966 households have 3 children, 208 households have 4 children, 17 households have 5 children, and 2 households have 6 children.
    \item \emph{Marital status}: The marital status categories include married, single, inferred married, inferred single, and unknown. They consist of 23834, 6165, 1146, 1792 and 4171 households, respectively. 
    \item \emph{Length of residence}: An integer ranging from 0 to 15. Its 25, 50 and 75 percentiles are 3, 7 and 14 respectively. A household has a value of 0 if it has been in residence for less than one year, and a value of 15 if it has been in residence for fifteen years or more. The values of 1 through 14 refer to the number of years the household has been in residence. The mean of this variable is equal to 7.76.\footnote{The raw data indicate that number of households in each of those categories are respectively, 1715,
 3512,
 3272,
 2816,
 2423,
 2470,
 2095,
 2100,
 1909,
 1688,
 1203,
 984,
 868,
 633,
 766,
 8654.}
    \item\emph{Number of cars}: The data set shows that 10778 households have 1 car, 5748 households have 2 cars, 2383 households have 3 cars, and the remaining households do not own a car.
    \item \emph{Number of credit cards}: This is an integer from 0 to 6. We observe from that the data set that 1193 households, 3006 households, 6240 households,  11693 households, 12983 households, 1958 households and 35 households have 0, 1, 2, 3, 4, 5, 6 credit cards respectively.
    \item \emph{Account types}: Cash (12618 accounts), Individual  Retirement (18734 accounts), Keogh (478 accounts), Margin (5462 accounts), Schwab One (16849 accounts).
    \begin{itemize}
        \item The \emph{Cash Account} is a type of brokerage account, in which the investor must pay the full amount for securities purchased, and he is not allowed to borrow funds from his broker to pay for transactions in the account.
        \item The \emph{Individual Retirement Account (IRA)} is an investment account for  retirement savings.
        \item The \emph{Keogh Account} is a tax-deferred pension plan available to self-employed individuals or unincorporated businesses for retirement purposes. 
        \item The \emph{Margin Account} is a brokerage account, in which the broker lends the customer cash to purchase stocks or other financial products. 
    \end{itemize}    
    \item \emph{Client segments}: Households can be grouped into three categories: active trader (6450 accounts), affluent household (10325 accounts), general brokerage (37366 accounts):
    \begin{itemize}
        \item \emph{Affluent household}: Households which hold more than \$100,000 in equity at any point in time.
        \item \emph{Active trader}:  Households which make more than 48 trades in any year. If a household qualifies as either active trader or affluent, it is assigned the active trader label.
        \item \emph{General brokerage}: All other households are labeled as general. 
    \end{itemize}
\end{enumerate} 
 
\subsection{Fama-French Industry Portfolios and CRSP Dataset}

We use data from Fama-French 49 industry portfolios to characterize the capital market line. The portfolio consists of all NYSE, AMEX, and NASDAQ stocks classified based on their Standard Industrial Classification (SIC) codes.\footnote{The SIC is a system used to classify industries based on a four-digit code. Established in the United States in 1937, it is used by government agencies to classify industry areas.} We use the value-weighted daily returns of the industry portfolio. In addition, we obtain the historical returns of stocks in the households' portfolios from the CRSP dataset. The historical returns of the 49 industry portfolios and investors' portfolios span a  period of three years.

\subsection{Fama-French Factor Dataset}

We use a Fama-French five-factor dataset to estimate the expected return and covariance matrix of the portfolios described above. We estimate the coefficients $(\mu, \Sigma)$, both for the 49-industry portfolio and investors' portfolios.

\cite{Fama_French:1993} developed an asset pricing model consisting of three factors. Their model is able to capture anomalies that are not explained by the capital asset pricing model developed by \cite{Sharpe:1964}. Roughly speaking, the excess return $(r_{i,t} - r_{f,t})$ of an asset $i$ at time $t$ linearly depends on the excess return of the market $(r_{mkt, t} - r_{f,t})$, the difference between the returns of a portfolio consisting of small cap and large cap stocks, denoted by $SMB_t$ at time $t$, and the difference between the returns of a portfolio composed of high book-to-market and low book-to-market ratio stocks, $HML_t$, at time $t$. In a subsequent study, \cite{Fama_French:2014} added the profitability and investment factors into the three-factor model, leading to a five-factor model. In addition to the aforementioned three factors, $RMW_t$ denotes the return spread of the most profitable over the least profitable firms, and $CMA_t$ denotes the return spread of firms that invest conservatively over those that invest aggressively.

\section{Estimation Strategy}\label{sec:implement}

The methodology discussed in Section \ref{sec:method} allows inferring investors' risk aversions and portfolio efficiencies from (i) the capital market line, and (ii) the mean and standard deviation of returns of each portfolio in each account.

First, we estimate the mean and covariance matrix of portfolio returns in each account for each household, as well as those of the 49-industry portfolio. The Fama-French five-factor model states that the excess return of  asset $i$ over the risk free rate, $(r_{i,t} - r_{f,t})$, in a universe of $p$ assets satisfies:
\begin{multline}
		r_{i,t} - r_{f, t} = \alpha_{i} + \beta_{i1} (r_{mkt, t} - r_{f,t}) + \beta_{i2} SMB_t + \beta_{i3} HML_t \\+ \beta_{i4} RMW_t + \beta_{i5} CMA_t + \e_{i,t},  
\end{multline}
where $\beta_{ij}, \ i = 1, \dots, p$ and $j = 1, \dots, 5$ are unknown factor loadings, and $\e_{1, \cdot}, \dots, \e_{p, \cdot}$ are $p$ idiosyncratic error terms. These errors have zero mean, and are uncorrelated with the specified factors $f$, where we set 
$$f := [(r_{mkt} - r_f), SMB, HML, RMW, CMA]^T.$$ Given the realizations of observed factors at $t = 1,...,T$, the factor betas for each asset are estimated from multivariate regression.  Thus, the estimated expected return vector $(\hat{\mu}^T)$ and the estimated covariance matrix ($\hat{\Sigma}$) are 
\begin{equation*}
		\hat{\mu}  = r_{f} + \hat{\alpha} + \hat{\beta} \bar{f},
\end{equation*}
\begin{equation*}
	\hat{\Sigma} = \hat{\beta} \widehat{\Cov}(f) \hat{\beta}^T + diag(\Var(\e_{1}), \dots, \Var(\e_{p})),
\end{equation*}
where $\hat{\beta}$ is the matrix of estimated regression coefficients, and $\widehat{\Cov}(f) $ is the sample covariance matrix of the factors $f$.

Next, the capital market line specifies the optimal combination of a risk-free asset and the market portfolio, and it is tangent to the efficient frontier. We characterize the market portfolio from Fama-French 49-industry portfolio, which assign all stocks from NYSE, AMEX, and NASDAQ to 49 categories based on their SIC codes.
Such a market portfolio can be computed explicitly, as we show in the next proposition.

\begin{prop}\label{prop}
Consider a market consists of $p$ risky assets, and let $r_f$ be the risk-free rate. Let  $\mu = [\mu_1, \mu_2, \dots, \mu_p]^T$ be a column vector of mean returns of $p$ risky assets, and $\Sigma$ the covariance matrix of returns. Then the expected return and standard deviation of the market portfolio admits the explicit expression given by
\begin{equation}
	\begin{aligned}
		\mu_{mkt} = & \frac{A - B r_f}{B - C r_f},\\
		\sigma_{mkt} = &  \frac{\sqrt{A - 2B r_f + C r_f^2}}{B -  C r_f},
	\end{aligned}
\end{equation}
where \begin{equation}
\begin{aligned}
 A = & \mu^T \Sigma^{-1} \mu,\\
 B = & \mu^T \Sigma^{-1} e = e^T \Sigma^{-1} \mu,\\
 C = & e^T \Sigma^{-1} e.
\end{aligned}
\end{equation}
\end{prop}

Using the above expressions for the expected return and volatility of the market portfolio, we can immediately obtain the capital market line, which connects the risk-free rate with the market portfolio.  
Hence, we can obtain the mean return and volatility for each portfolio, at any time, from the Fama-French five-factor model. The capital market line is also explicitly given from the above computation. The risk aversion coefficients are readily obtained by projecting the observed portfolios on the capital market line according to the method described in Section \ref{sec:method}.

\section{Descriptive Statistics of Risk Preferences and Portfolio Efficiency}\label{sec:hist}

We imply the risk aversion coefficient for each account of each household. This yields a total of 54,141 accounts, together with 1,981,277 risk aversion coefficients, one for each of the 72 monthly periods between 1991 and 1996 over which the household invests. The 25th, 50th, and 75th  percentiles of such risk aversion coefficients are 40.97,  67.49, and 108.74, respectively. 

\begin{figure}[htbp]
\begin{center}
  \includegraphics[width = 0.45\textwidth, height = 5cm]{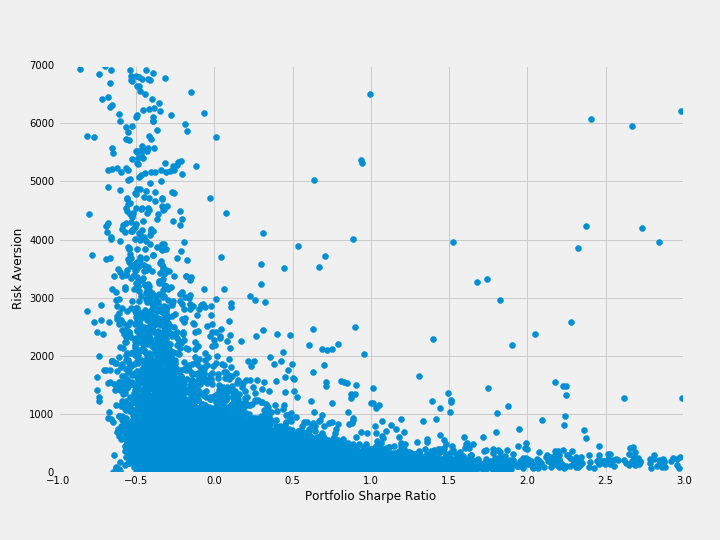}
    \includegraphics[width = 0.45\textwidth, height = 5cm]{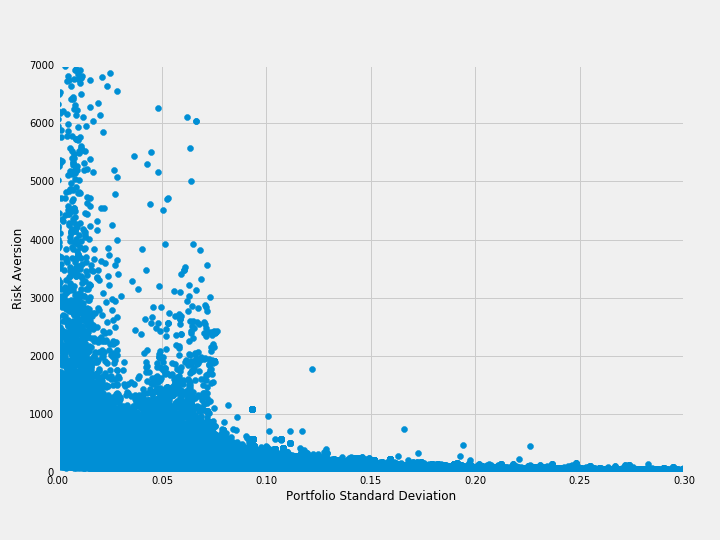}
  \caption{Left: Scatter plot of risk aversion versus portfolio Sharpe ratio. Right: Scatter plot of risk aversion versus portfolio standard deviation. }
  \label{fig:sr_sd}
    \end{center}
\end{figure}

Figure \ref{fig:sr_sd} presents scatter plots of risk aversion coefficients versus portfolios' Sharpe ratios and standard deviations. The data indicate that an investor becomes less risk averse as her portfolio Sharpe ratio increases, and that she becomes more risk averse if she experiences a loss and her portfolio Sharpe ratio becomes negative. {In summary, we find that investors who achieve higher portfolio Sharpe ratio in the stock market are willing to take more risk}. To the best of our knowledge, this finding has not been highlighted in earlier literature. Additionally, such finding is consistent with results in the behavioral finance literature. For instance, \cite{Thaler:1990} find that prior gains can increase  subjects' willingness to accept gambles, while prior losses  decrease their willingness to take risks. Moreover, and consistent with intuition, investors who are {willing to take more risk}, i.e., less risk averse, tend to hold more volatile portfolios. These findings are consistent with earlier studies, such as \cite{Cohn:1974}, who  elicits investors' risk preference information via questionnaires, \cite{Riley:1992} who uses financial data for a large random sample of U.S. households, and \cite{Bucciol_Miniaci:2011} who uses a survey of consumer finances dataset.

\begin{figure}[htbp]
\begin{center}
      \includegraphics[width = 0.45\textwidth, height = 5cm]{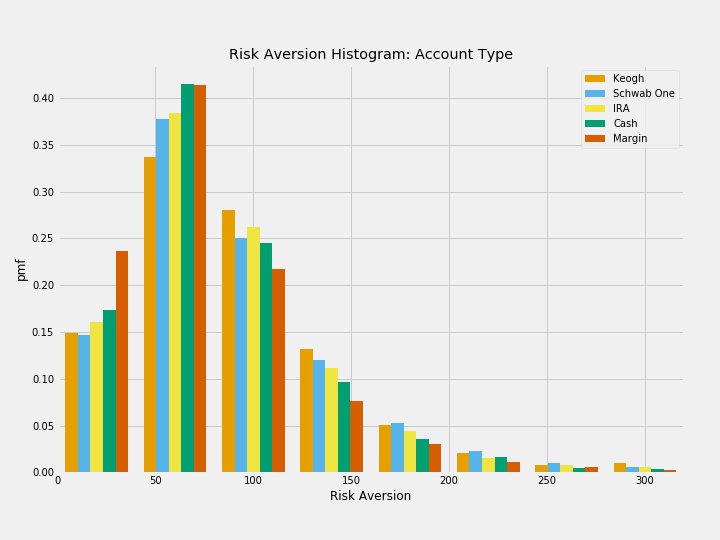}
      \quad \quad \quad 
    \includegraphics[width = 0.45\textwidth, height = 5cm]{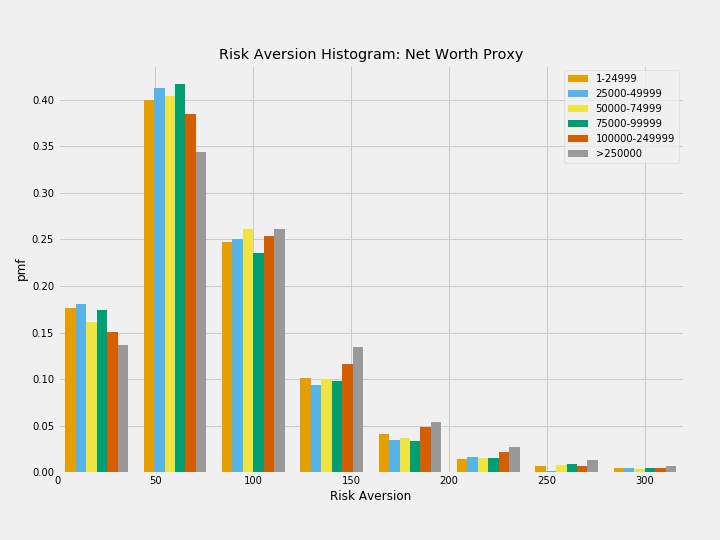}\\
    \vspace{1cm}
    \includegraphics[width = 0.45\textwidth, height = 5cm]{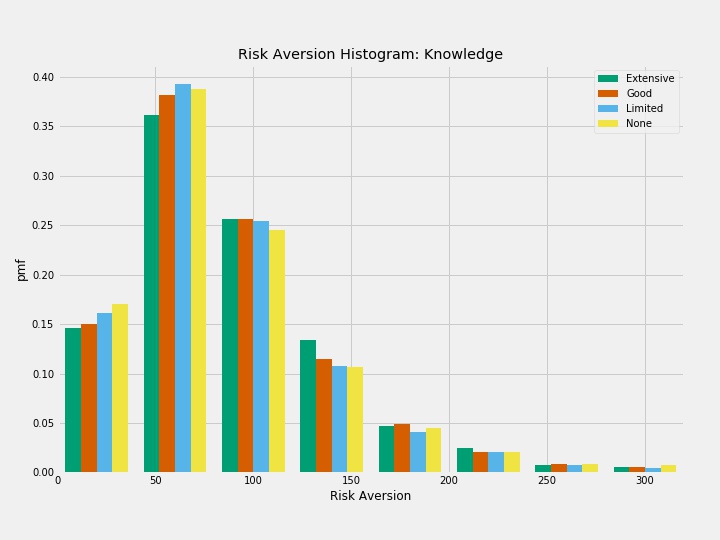}
    \quad \quad \quad 
    \includegraphics[width = 0.45\textwidth, height = 5cm]{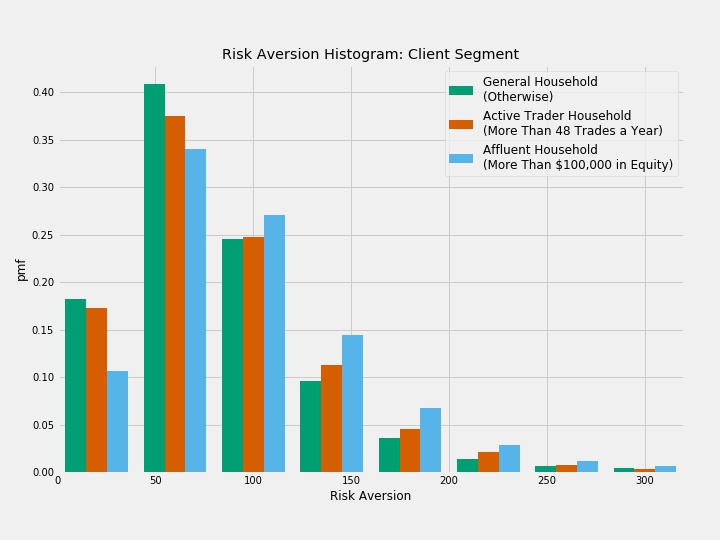}\\
        \vspace{1cm}
  \includegraphics[width = 0.45\textwidth, height = 5cm]{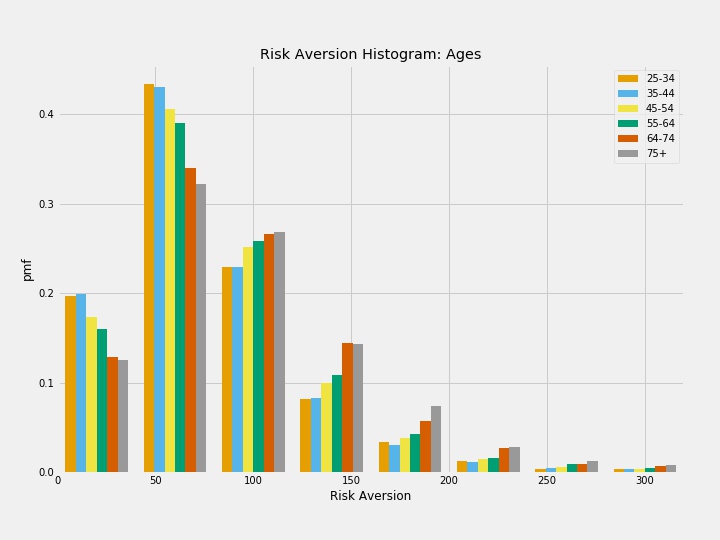}
  \quad \quad \quad 
    \includegraphics[width = 0.45\textwidth, height = 5cm]{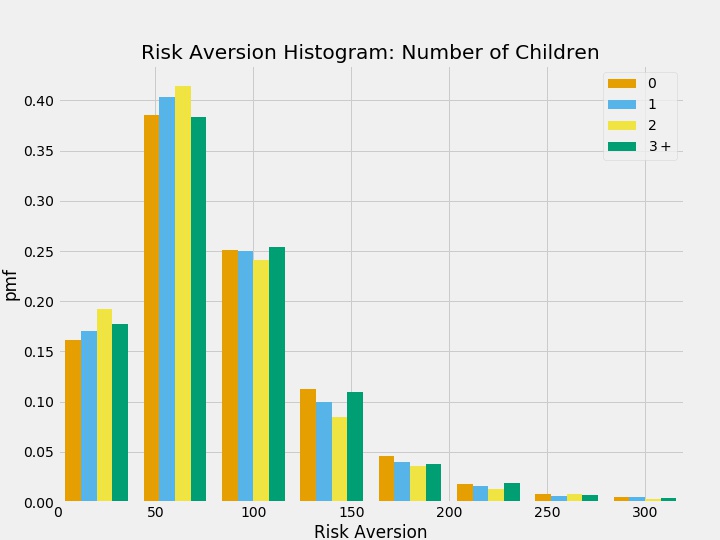}
  \caption{Histograms of implied risk aversion coefficients based on account types, net worth, financial knowledge, client segments, ages, and number of children of households. } 
  \label{fig:rv}
    \end{center}
\end{figure}

Next, we analyze the dependence of risk preferences on investors' demographics, financial literacy and wealth characteristics.  Figure \ref{fig:rv} presents histograms of the time average of investors' risk aversion coefficients,  obtained from the procedures discussed in Sections \ref{sec:method} and \ref{sec:implement}, versus investors' account types, net worth, knowledge, segments, ages, and number of children in the household.

The first plot in Figure \ref{fig:rv} presents the distribution of risk preferences across all account types. The data show that more than 60\% of the margin accounts have risk aversion smaller than 80, which is the highest percentage across all  account types. 
We also observe that more than 50\% of investors with Keogh account have risk aversion greater than 80, which is the highest risk aversion among account types. The other types of accounts all lie somewhere in the middle. These observed statistics can be understood from the purposes of these accounts: the margin account allows customers to borrow money from the broker, while the Keogh Account is for retirement purposes only and thus associated with a high level of risk aversion.

The second and the third graph in Figure \ref{fig:rv} show that more wealth and financially more literate investors tend to be more risk averse. This can be understood by the fact that financially more knowledgeable and wealth investors are able to diversify their risks better by holding more securities.

The fourth plot of Figure \ref{fig:rv} shows that  general households are the least risk averse, while affluent households are the most risk averse. This can be explained by the fact that affluent households hold more diversified portfolios, which imply a  higher risk aversion. Furthermore, the fifth plot of Figure \ref{fig:rv} indicates that, investors become more risk averse as they get older. 
This finding is consistent with a large body of literature (see, for instance, \cite{Sahm:2012}, \cite{Bucciol_Miniaci:2011}, \cite{Palsson:1996}, and \cite{Morin_Suarez:1983}). Lastly, investors tend to be less risk averse as the number of children grows from 0 to 2, but investors with more than 3 children tend to be more risk averse. The ``number of children'' factor might have little impact on investors' risk aversion, since the distribution of risk aversion across the number of children is not very obvious.

\begin{figure}[htbp]
\begin{center}
    \includegraphics[width = 0.45\textwidth, height = 5cm]{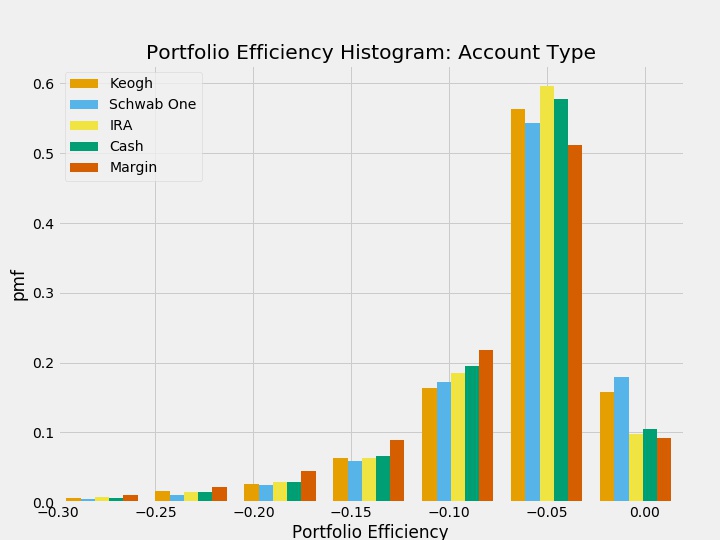}
          \quad \quad \quad 
    \includegraphics[width = 0.45\textwidth, height = 5cm]{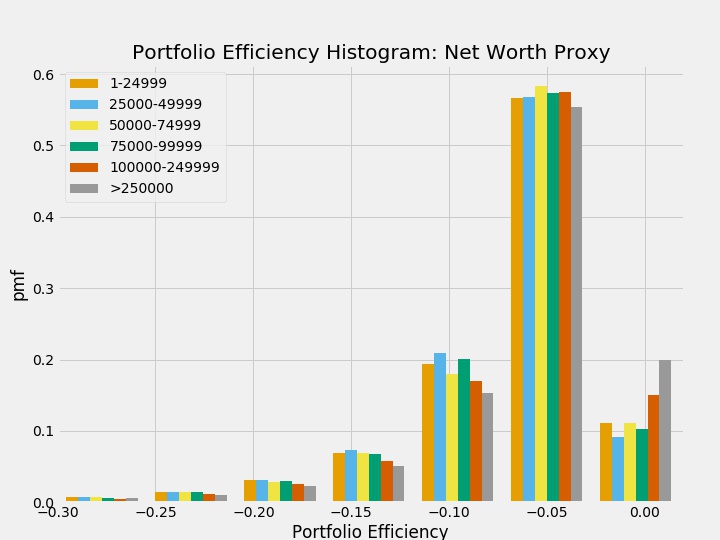}\\
            \vspace{1cm}
    \includegraphics[width = 0.45\textwidth, height = 5cm]{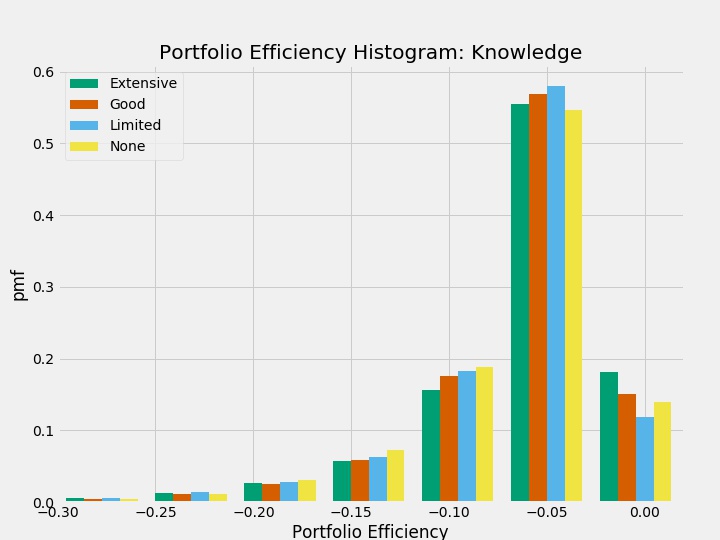}
          \quad \quad \quad 
    \includegraphics[width = 0.45\textwidth, height = 5cm]{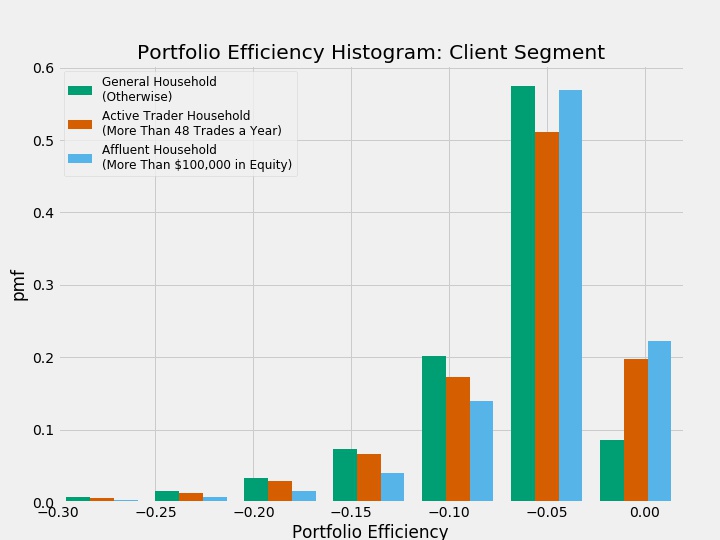}\\
            \vspace{1cm}
    \includegraphics[width = 0.45\textwidth, height = 5cm]{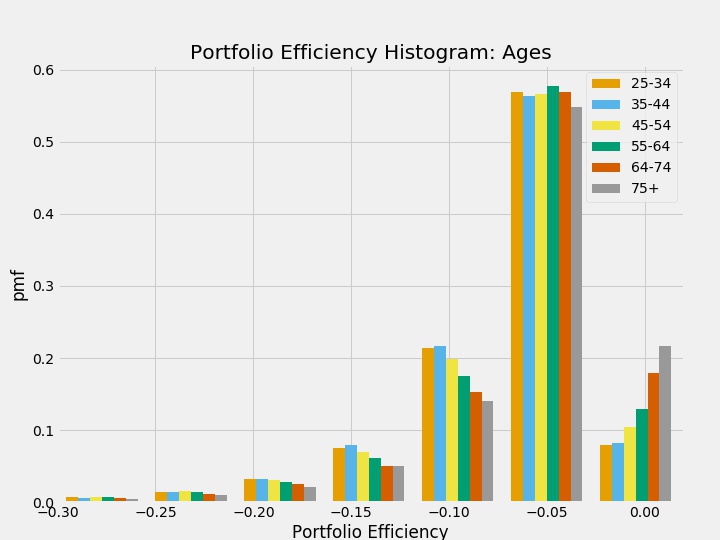}
          \quad \quad \quad 
    \includegraphics[width = 0.45\textwidth, height = 5cm]{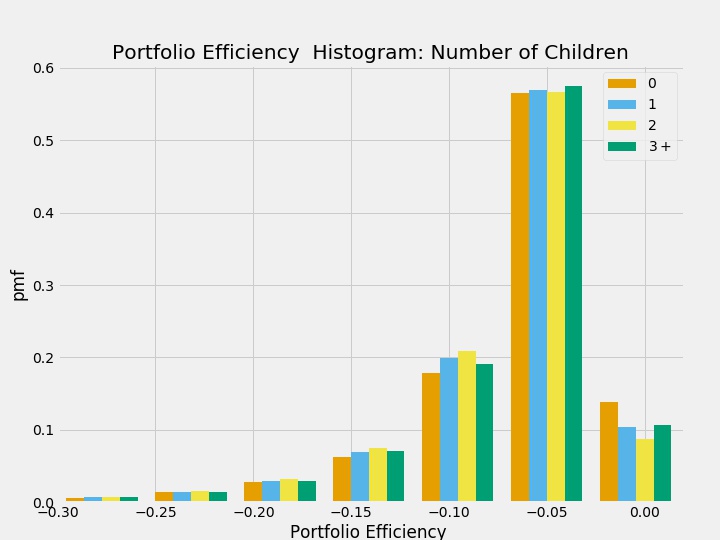}
  \caption{Histograms of portfolio efficiency based on account types, net worth, financial knowledge, client segments, ages, and number of children of households.} 
  \label{fig:eff}
    \end{center}
\end{figure}

We conclude with an investigation of how investors' portfolio efficiency depends on demographics and account information. The 25\%, 50\% and 75\% percentiles of the portfolio efficiencies are -0.073, -0.05, and -0.035 respectively. Figure \ref{fig:eff} reports histograms of investors' portfolio efficiency versus investors' account types, net worth, knowledge, segments, ages, and the number of children in their households. The top left histogram shows that portfolios in margin and cash accounts are less efficient than those in the retirement related accounts. As for the the relationship between portfolio efficiency and net worth, from the second plot in Figure \ref{fig:eff}, we see that investors with more net worth hold more efficient portfolios when the net worth exceeds \$75000.  The third histogram indicates that financially more literate investors hold more efficient portfolios. The rationale here is that investors with higher financial literacy are better at diversifying portfolios (see also \cite{Abreu:2010} and \cite{Guiso:2008}). {Our analysis shows that, on average, investors with limited, good, and expert financial knowledge hold, respectively, 2.69, 3.22, and 3.67 stocks.} 
Moreover, affluent traders' portfolios are the most efficient, while general households' portfolios are the least efficient. 
The fifth histogram shows that investors hold more efficient portfolios as they become older, which can be explained by a higher financial experience gained throughout the years. Lastly, investors' portfolios are less efficient if the number of children in the investor households grows from 0 to 2.

\section{Regression Analysis}\label{sec:regression}

In this section, we examine the determinants of risk aversion coefficients and portfolio efficiencies using a linear panel regression. We also compare our results to existing literature.

\subsection{Determinants of Risk Aversion}
We perform a linear regression of implied risk preferences on investors' demographic information, account information, portfolio's characteristics, and market conditions. The corresponding results are reported in Table \ref{table:plm_rv}. 

We first remove  outliers, i.e., implied risk aversions that are above or below 1.5 times the interquartile range.   After removing 119,315 outliers out of 1,861,962 points, the adjusted 25\%, 50\% and 75\% percentiles of the implied risk aversions are  39.61, 64.14, and 99.01. The demographic information includes net worth proxy, income proxy, financial literacy, ages, number of children, marital status, length of residence, number of cars, and number of credit cards. The account information includes investors' account types, client segments, and investors' portfolio information include the number of stocks in their portfolio, portfolio Sharpe ratio, portfolio expected return, and portfolio volatility. Lastly, we use the VIX index as a proxy for market volatility. The VIX typically rises during times of financial stress and market selloffs, and falls as investors become complacent.

Table \ref{table:plm_rv} reports the results of the linear panel regression model. We account for time fixed effects in Column (1) and Column (2) of Table \ref{table:plm_rv}, where we control for variables that are constant across entities but vary over time, such as market volatility. 

Column (1) shows the existence of a statistically significant positive relation between investors' implied risk aversion and their net worth. Investors' financial literacy is also a statistically significant predictor. Investors who are financially less literate, i.e., whose knowledge moves from extensive to limited, are less risk averse. This confirms statistically the visual findings from Figure \ref{fig:rv}. {It is worth remarking that existing literature (e.g. \cite{Riley:1992}, \cite{Guiso:2008}, and \cite{Rooij:2011}) has measured the individual risk preferences of an investor from its proportion of total wealth allocated to risky assets. They argue that, with an increase in wealth and education level, the investors' willingness to accept more risk would increase. This means that investors are willing to allocate a larger proportion of total wealth to the risky assets. We would like to remark that our results are not contradicting theirs. In the market of stocks we are considering, wealthier and financially more literate investors are more likely to hold a more diversified portfolio, relative to less wealthy and financially less savy investors. This in turn reduces the risk they are taking, and as a result their implied risk aversion coefficient is higher.}

Interestingly, our regression shows a statistically significant and negative linear dependence between risk aversion and investors' income, for large incomes (greater than \$75,000 per year.) The higher willingness to take risk from investors with larger income is consistent with the findings of \cite{Dohmen:2010}. Furthermore, 
it is evident from  Column (1) of Table \ref{table:plm_rv} that, compared to the Cash account, households are most risk averse on accounts related to retirement, while they are most risk tolerant on Margin accounts. Furthermore, affluent households who are wealthy tend to be the most risk averse investors. These findings confirm statistically the visual intuition gained from Figure \ref{fig:rv}. In addition, our regressions also indicate that investors become more risk averse as they get older, stay in the same residence for a longer time, and have a higher number of cars, while they get less risk averse if they hold more credit cards.
\footnote{We remark that the data set in \cite{Guiso:2008} consists of only 2377 observations, and he finds a $R^2$ coefficient of about 12\%. Our data set consists of a significantly large number of observations, $1,742,647$, and we find a $R2$ square of $3.1\%$}

After controlling for the demographic information and account information, in  Column (2)  of Table \ref{table:plm_rv},  we confirm statistically that investors holding more securities portfolios are more risk averse, because a diversified portfolio  reduces idiosyncratic risk. However, we also find that the impact of holding a larger number of stocks on the investor's risk preferences strongly depend on the client segment. Investors become less risk averse, as their portfolio Sharpe ratio, expected return and portfolio standard deviation increase. This finding is intuitive: an investor has to take more risk to achieve a larger payoff, either in terms of expected portfolio return or portfolio Sharpe ratio. 
The proportion of explained variation if one accounts for investors' demographics, account information, and portfolio characteristics is 41.9\%.

In Column (3) and Column (4) of Table \ref{table:plm_rv}, we analyze the relation between market and portfolio characteristics and investors' risk preferences via a panel regression that accounts for fixed entity effects.\footnote{This is obtained by removing the time-invariant characteristics, such as demographic and account information.} Column (3) of Table \ref{table:plm_rv} shows that market variables impact risk preferences, and the $R^2$ is equal to 0.039. Specifically, investors become more risk averse when the aggregate risk in the market is high. \cite{Malmendier:2011} come to a similar conclusion by measuring risk attitudes using four measures (survey of willingness to take risk, stock market participation, bond market participation, and the  fraction of liquid assets invested in stocks) and data from Survey  of Consumer Finances from 1960 to 2007. If we additionally control for portfolio characteristics, such as number of stocks, portfolio Sharpe ratio, portfolio mean and volatility beyond the VIX index, the proportion of explained variation increases to nearly $10\%$.

\subsection{Panel Regressions on Portfolio Efficiency}

Despite investors' net worth are significant, there seems to be no obvious linear relationship between net worth and their portfolio efficiency (see Column (1) of Table \ref{table:plm_eff}).  Portfolios are more efficient for investors whose net worth is larger than \$75000 per year.\footnote{The comparison of the coefficients of the categorical variables are based on the reference group. The reference group in this regression consists of investors whose age is between 18 and 24, and who are inferred married, and affluent traders invested in Cash account with 0-24999 net worth, 0-24999 annual income, and extensive financial knowledge.} In addition, investors who are financially less literate also have less efficient  portfolios. These findings are broadly consistent with earlier studies (see \cite{Calvet:2009} and \cite{Calvet:2007}), who also find that investors with greater financial sophistication, as measured for instance by wealth or education, tend to commit less financial mistakes, {such as underdiversification, risky share inertia, sale of winning stocks and holding of losing stocks} We also find that annual income has a negative impact on portfolio efficiency.
The regressions indicate a positive relation between investors' age and portfolios efficiency. Because financial knowledge and experience typically increase with age, this provides an explanation for why older investors hold more efficient portfolios. 
Column (1) in Table \ref{table:plm_eff} also shows that investors' portfolios becomes less efficient as the number of children grows, and the number of their credit cards increases.  Though many demographic variables are significant, the percentage of variability explained remains low ($R^2= 0.024$). 

In Column (2) of Table \ref{table:plm_eff}, we control for investors' account information, i.e., their account types and segments. We find that portfolios in Margin and Cash accounts are less efficient than those in retirement related accounts. This can be understood in terms of the fact that investments in retirement related accounts are less aggressive, and less volatile.
In addition,  as compared to affluent traders, both general and active trades' portfolios are less efficient.  General traders, who are the least experienced, are more likely to hold the least efficient portfolios. Controlling for account information in the panel regression increases the percentage of explained variation from 0.024 to 0.041.

In addition to demographics and account information, we also control for portfolio variables in Column (3) of Table \ref{table:plm_eff}.  First, we observe that portfolios become more efficient as the number of stocks invested increases. This is consistent with intuition: increasing the number of stocks yields a higher portfolio diversification, and it is well known from Markowitz theory that all efficient portfolios are well-diversified. 
Moreover, portfolio Sharpe ratio and mean return positively impact efficiency, while the efficiency of a portfolio decreases as the portfolio becomes more volatile. The Sharpe ratio measures the amount of return per unit of risk. The positive relationship between portfolio efficiency and  portfolio's Sharpe ratio is intuitive.\footnote{\cite{Gibbons:1989} proposed a test of efficiency based on the based on the Sharpe ratios difference between the observed and the efficient portfolio.} Moreover, any rational investor would prefer to maximize return for a specified level of risk, or equivalently, to minimize risk for a given level of return. This explains why  expected return impacts positively, while volatility impacts negatively, the portfolio's efficiency. Controlling for demographics, account, and portfolio characteristics altogether explain all variation in the data, i.e., the $R^2$ reaches 99.4\%.

We next turn to analyze the impact of market conditions on portfolio efficiency, accounting for fixed entity effects. 
Column (4) of Table \ref{table:plm_eff} shows that the VIX index has a negative impact on portfolio efficiency. However, there appears to be no evidence of any correlation between these two variables ($R^2$ is almost 0, and adjusted $R^2$ is negative). 
Accounting for portfolio characteristics, in addition to  market variables, the $R^2$ coefficient reaches $91.5\%$ (see Column (4) of Table \ref{table:plm_eff}). Hence, we can conclude that the portfolio efficiency is primarily driven by portfolio characteristics and very little by market variables.

Lastly, to further investigate which portfolio variables contribute the most to portfolio efficiency, we perform two way fixed effect regressions on portfolio variables in Table \ref{table:eff_twoway}. We then conclude that portfolio volatility is a driving factor of portfolio efficiency. This suggests that, as investors' portfolios become more volatile, their portfolios also become less efficient, and portfolio under-diversification could be one of the reasons for this decrease in efficiency.

\section{Conclusions}\label{sec:conclusions}
In this study, we have analyzed portfolio stock holdings of investors and related the implied risk aversion profiles to their demographics. Our study shows that the majority of investors' portfolios are inefficient in the sense that they lie below the capital market line. As the capital market line traces the optimal portfolios associated with any possible risk aversion parameter in a mean-variance portfolio criterion, we use it as a baseline to infer the efficiency of their portfolios.

We propose an approach to assess an investor's risk preference from her investment portfolio in a mean-variance Markowitz context, and obtained a simple closed-form expression for the implied risk-aversion coefficient. Our formula shows that investors are less risk averse if they hold more volatile portfolios or portfolios with higher Sharpe ratio. We have measured the portfolio efficiency as the Euclidean distance between the observed portfolio and the projected portfolio on the capital market line. The latter is the closest optimal portfolio an investor can attain by eliminating excess volatility and increasing the expected return while maintaining the same risk preference. 

We have used the derived formulas on a dataset of end of month trading records of investors, along with the corresponding demographic information, and self-assessed financial knowledge.
Our statistical analysis finds a positive correlation between investors' implied risk aversion and their net worth, financial literacy, ages, and number of invested securities. Investors become less risk averse as their portfolio Sharpe ratio, expected return and portfolio standard deviation increases. We also find a positive correlation between investor's implied risk aversion and aggregate risk, i.e., the investor becomes more risk averse if the VIX increases. {Our statistical analysis shows that wealthier, older and financially more literate investors hold more efficient portfolios, and that more diversification in terms of number of invested stocks, higher portfolio Sharpe ratio and mean return, also improve the efficiency of the portfolio}. Finally, we find that volatility captures significant variation of portfolio's efficiency, and is negatively correlated with portfolio efficiency.

\pagebreak

\setlength\LTleft{-0.5in}
\setlength\LTright{-1in plus 1 fill}
\tiny
\begin{longtable}{@{\extracolsep{\fill}}lcccc} 
   \caption{Linear Panel Regression Models of Risk Aversion \label{table:plm_rv}} \\
\\[-1.8ex]\hline 
\hline \\[-1.8ex] 
 & \multicolumn{4}{c}{\textit{Dependent variable:}} \\ 
\cline{2-5} 
\\[-1.8ex] & \multicolumn{4}{c}{Risk Aversion} \\ 
  & (1) & (2) & (3) & (4) \\ 
\hline \\[-1.8ex] 
 Net Worth 25,000-49,999 & $-$0.332$^{*}$ & 0.075 &  &  \\ 
  & & & & \\ 
 Net Worth 50,000-74,999 & 1.227$^{***}$ & 1.337$^{***}$ &  &  \\ 
  & & & & \\ 
 Net Worth 75,000-99,999 & $-$0.38$^{*}$ & $-$0.102 &  &  \\ 
  & & & & \\ 
 Net Worth 100,000-249,999 & 2.242$^{***}$ & 1.38$^{***}$ &  &  \\ 
  & & & & \\ 
 Net Worth $>$250,000 & 4.267$^{***}$ & 2.677$^{***}$ &  &  \\ 
  & & & & \\ 
 Income 25,000-49,999 & 2.165e-03 & 0.177 &  &  \\ 
  & & & & \\ 
 Income 50,000-74,999 & 0.061 & $-$4.550e-05 &  &  \\ 
  & & & & \\ 
 Income 75,000-99,999 & $-$0.945$^{***}$ & $-$0.664$^{***}$ &  &  \\ 
  & & & & \\ 
 Income $>$100,000 & $-$2.12$^{***}$ & $-$1.388$^{***}$ &  &  \\ 
  & & & & \\ 
 Good Knowledge & $-$1.395$^{***}$ & $-$1.106$^{***}$ &  &  \\ 
  & & & & \\ 
 Limited Knowledge & $-$2.772$^{***}$ & $-$1.263$^{***}$ &  &  \\ 
  & & & & \\ 
 None Knowledge & $-$1.655$^{***}$ & $-$1.256$^{***}$ &  &  \\ 
  & & & & \\ 
 Unknown Knowledge & $-$1.703$^{***}$ & $-$1.002$^{***}$ &  &  \\ 
  & & & & \\ 
 Ages 25-34 & $-$1.072 & 0.444 &  &  \\ 
  & & & & \\ 
 Ages 35-44 & $-$1.475$^{*}$ & $-$0.149 &  &  \\ 
  & & & & \\ 
 Ages 45-54 & 1.067 & 1.116$^{*}$ &  &  \\ 
  & & & & \\ 
 Ages 55-64 & 2.796$^{***}$ & 2.131$^{***}$ &  &  \\ 
  & & & & \\ 
 Ages 67-74 & 8.402$^{***}$ & 5.255$^{***}$ &  &  \\ 
  & & & & \\ 
 Ages $>$75 & 11.168$^{***}$ & 6.626$^{***}$ &  &  \\ 
  & & & & \\ 
 Num. of Children & $-$0.033 & 0.034 &  &  \\ 
  & & & & \\ 
 Inferred Single & $-$0.749$^{***}$ & $-$0.936$^{***}$ &  &  \\ 
  & & & & \\ 
 Married & $-$0.333$^{**}$ & $-$0.34$^{***}$ &  &  \\ 
  & & & & \\ 
 Single & $-$0.168 & $-$0.844$^{***}$ &  &  \\ 
  & & & & \\ 
 Unknown Marital & $-$0.482$^{***}$ & $-$0.629$^{***}$ &  &  \\ 
  & & & & \\ 
 Length of Residence & 0.062$^{***}$ & $-$6.029e-03 &  &  \\ 
  & & & & \\ 
 Num. of Cars & 0.248$^{***}$ & 0.154$^{***}$ &  &  \\ 
  & & & & \\ 
 Num. of Credit Cards & $-$0.649$^{***}$ & $-$0.311$^{***}$ &  &  \\ 
  & & & & \\ 
 IRA Account & 2.681$^{***}$ & 1.796$^{***}$ &  &  \\ 
  & & & & \\ 
 Keogh Account & 4.428$^{***}$ & 1.569$^{***}$ &  &  \\ 
  & & & & \\ 
 Margin Account & $-$4.424$^{***}$ & $-$4.225$^{***}$ &  &  \\ 
  & & & & \\ 
 Schwab Account & 4.431$^{***}$ & $-$0.039 &  &  \\ 
  & & & & \\ 
 General Brokerage & $-$8.924$^{***}$ & $-$4.679$^{***}$ &  &  \\ 
  & & & & \\ 
 Active Trader & $-$7.664$^{***}$ & $-$0.995$^{***}$ &  &  \\ 
  & & & & \\ 
 Num. of Stocks &  & 4.134$^{***}$ &  & 3.639$^{***}$ \\ 
  & & & & \\ 
 Portfolio Sharpe Ratio &  & $-$22.829$^{***}$ &  & $-$8.614$^{***}$ \\ 
  & & & & \\ 
 Portfolio Expected Return &  & $-$411.686$^{***}$ &  & $-$188.991$^{***}$ \\ 
  & & & & \\ 
 Portfolio Std. Deviation &  & $-$145.647$^{***}$ &  & $-$65.647$^{***}$ \\ 
  & & & & \\ 
 General Brokerage x Num. of Stocks &  & 0.68$^{***}$ &  & 0.531$^{***}$ \\ 
  & & & & \\ 
 Active Trader x Num. of Stocks &  & $-$1.653$^{***}$ &  & $-$1.387$^{***}$ \\ 
  & & & & \\ 
 VIX Index &  &  & 2.572$^{***}$ & 2.632$^{***}$ \\ 
  & & & & \\ 
\hline \\[-1.8ex] 
Observations & 1861962 & 1861962 & 1861962 & 1861962 \\ 
R$^{2}$ & 0.031 & 0.419 & 0.039 & 0.096 \\ 
Adjusted R$^{2}$ & 0.031 & 0.419 & 0.01 & 0.069 \\ 
F Statistic & 1.828e+03$^{***}$  & 3.447e+04$^{***}$  & 7.358e+04$^{***}$  & 2.755e+04$^{***}$  \\ 
& (df = 33; 1861858) & (df = 39; 1861852) & (df = 1; 1807948) & \\
\hline 
\hline \\[-1.8ex] 
\textit{Note:}  & \multicolumn{4}{r}{$^{*}$p$<$0.1; $^{**}$p$<$0.05; $^{***}$p$<$0.01} \\ 
\end{longtable}  
\normalsize
\pagebreak

\begin{landscape}
\pagestyle{empty}
\setlength\LTleft{0.1in}
\setlength\LTright{-1in plus 1 fill}
\tiny
\begin{longtable}{@{\extracolsep{\fill}}lcccccc} 
\caption{Panel Regression Models of Portfolio Efficiency \label{table:plm_eff}} \\
\\[-1.8ex]\hline 
\hline \\[-1.8ex] 
 & \multicolumn{6}{c}{\textit{Dependent variable:}} \\ 
\cline{2-7} 
\\[-1.8ex] & \multicolumn{6}{c}{Portfolio Efficiency} \\ 
  & (1) & (2) & (3) & (4) & (5) \\ 
\hline \\[-1.8ex] 
 Net Worth 25,000-49,999 & $-$3.849e-04$^{**}$ & $-$3.922e-04$^{***}$ & $-$3.080e-06 &  &  \\ 
  & (1.511e-04) & (1.498e-04) & (1.170e-05) &  &  \\ 
  & & & & & \\ 
 Net Worth 50,000-74,999  & 2.493e-03$^{***}$ & 2.183e-03$^{***}$ & 2.017e-05 &  &  \\ 
  & (1.705e-04) & (1.690e-04) & (1.320e-05) &  &  \\ 
  & & & & & \\ 
 Net Worth 75,000-99,999 & 1.022e-03$^{***}$ & 6.513e-04$^{***}$ & 7.030e-06 &  &  \\ 
  & (1.625e-04) & (1.611e-04) & (1.258e-05) &  &  \\ 
  & & & & & \\ 
 Net Worth 100,000-249,999 & 4.407e-03$^{***}$ & 2.784e-03$^{***}$ & 1.207e-05 &  &  \\ 
  & (1.365e-04) & (1.356e-04) & (1.059e-05) &  &  \\ 
  & & & & & \\ 
 Net Worth $>$250,000 & 6.879e-03$^{***}$ & 4.331e-03$^{***}$ & 1.754e-05 &  &  \\ 
  & (1.438e-04) & (1.434e-04) & (1.120e-05) &  &  \\ 
  & & & & & \\ 
 Income 25,000-49,999 & $-$1.143e-03$^{***}$ & $-$7.723e-04$^{***}$ & $-$1.076e-05 &  &  \\ 
  & (1.448e-04) & (1.436e-04) & (1.121e-05) &  &  \\ 
  & & & & & \\ 
 Income 50,000-74,999 & $-$9.341e-04$^{***}$ & $-$6.508e-04$^{***}$ & $-$2.167e-05$^{*}$ &  &  \\ 
  & (1.533e-04) & (1.520e-04) & (1.187e-05) &  &  \\ 
  & & & & & \\ 
 Income 75,000-99,999 & $-$1.614e-03$^{***}$ & $-$1.431e-03$^{***}$ & $-$2.736e-05$^{**}$ &  &  \\ 
  & (1.525e-04) & (1.512e-04) & (1.181e-05) &  &  \\ 
  & & & & & \\ 
 Income $>$100,000 & $-$1.829e-03$^{***}$ & $-$2.112e-03$^{***}$ & $-$1.560e-06 &  &  \\ 
  & (1.535e-04) & (1.522e-04) & (1.189e-05) &  &  \\ 
  & & & & & \\ 
 Good Knowledge & $-$1.140e-03$^{***}$ & $-$8.974e-04$^{***}$ & $-$1.678e-05$^{**}$ &  &  \\ 
  & (1.077e-04) & (1.070e-04) & (8.360e-06) &  &  \\ 
  & & & & & \\ 
 Limited Knowledge & $-$2.193e-03$^{***}$ & $-$1.527e-03$^{***}$ & $-$7.930e-06 &  &  \\ 
  & (1.169e-04) & (1.165e-04) & (9.100e-06) &  &  \\ 
  & & & & & \\ 
 None Knowledge & $-$2.559e-03$^{***}$ & $-$1.669e-03$^{***}$ & $-$1.020e-05 &  &  \\ 
  & (1.516e-04) & (1.506e-04) & (1.176e-05) &  &  \\ 
  & & & & & \\ 
 Unknown Knowledge & $-$2.313e-03$^{***}$ & $-$9.604e-04$^{***}$ & $-$6.120e-06 &  &  \\ 
  & (1.357e-04) & (1.353e-04) & (1.056e-05) &  &  \\ 
  & & & & & \\ 
 Ages 25-34 & 1.981e-03$^{***}$ & 2.158e-03$^{***}$ & $-$6.650e-05 &  &  \\ 
  & (6.801e-04) & (6.742e-04) & (5.264e-05) &  &  \\ 
  & & & & & \\ 
 Ages 35-44 & 1.258e-03$^{*}$ & 1.143e-03$^{*}$ & $-$4.781e-05 &  &  \\ 
  & (6.734e-04) & (6.675e-04) & (5.212e-05) &  &  \\ 
  & & & & & \\ 
 Ages 45-54 & 3.705e-03$^{***}$ & 3.160e-03$^{***}$ & $-$5.289e-05 &  &  \\ 
  & (6.732e-04) & (6.673e-04) & (5.211e-05) &  &  \\ 
  & & & & & \\ 
 Ages 55-64 & 6.390e-03$^{***}$ & 5.643e-03$^{***}$ & $-$3.816e-05 &  &  \\ 
  & (6.740e-04) & (6.682e-04) & (5.218e-05) &  &  \\ 
  & & & & & \\ 
 Ages 67-74 & 0.011$^{***}$ & 9.172e-03$^{***}$ & $-$2.266e-05 &  &  \\ 
  & (6.745e-04) & (6.687e-04) & (5.222e-05) &  &  \\ 
  & & & & & \\ 
 Ages $>$75 & 0.012$^{***}$ & 0.011$^{***}$ & $-$2.489e-05 &  &  \\ 
  & (6.755e-04) & (6.697e-04) & (5.229e-05) &  &  \\ 
  & & & & & \\ 
 Num. of Children & $-$4.664e-04$^{***}$ & $-$3.587e-04$^{***}$ & $-$2.450e-06 &  &  \\ 
  & (2.872e-05) & (2.849e-05) & (2.220e-06) &  &  \\ 
  & & & & & \\ 
 Inferred Single & $-$4.394e-04$^{***}$ & $-$4.887e-04$^{***}$ & $-$1.620e-05 &  &  \\ 
  & (1.644e-04) & (1.630e-04) & (1.273e-05) &  &  \\ 
  & & & & & \\ 
 Married  & $-$3.523e-05 & $-$2.863e-04$^{**}$ & 3.510e-06 &  &  \\ 
  & (1.273e-04) & (1.262e-04) & (9.860e-06) &  &  \\ 
  & & & & & \\ 
 Single & $-$1.802e-05 & $-$2.755e-04$^{**}$ & $-$1.315e-05 &  &  \\ 
  & (1.361e-04) & (1.350e-04) & (1.054e-05) &  &  \\ 
  & & & & & \\ 
 Unknown Marital & $-$5.063e-04$^{***}$ & $-$7.751e-04$^{***}$ & $-$8.340e-06 &  &  \\ 
  & (1.407e-04) & (1.395e-04) & (1.089e-05) &  &  \\ 
  & & & & & \\ 
 Length of Residence & 6.750e-06 & 7.100e-07 & $-$9.200e-07$^{***}$ &  &  \\ 
  & (4.460e-06) & (4.420e-06) & (3.500e-07) &  &  \\ 
  & & & & & \\ 
 Num. of Cars & 1.764e-05 & 2.536e-04$^{***}$ & 3.750e-06$^{**}$ &  &  \\ 
  & (2.371e-05) & (2.355e-05) & (1.840e-06) &  &  \\ 
  & & & & & \\ 
 Num. of Credit Cards & $-$2.610e-04$^{***}$ & $-$2.666e-04$^{***}$ & $-$4.510e-06$^{***}$ &  &  \\ 
  & (1.998e-05) & (1.982e-05) & (1.550e-06) &  &  \\ 
  & & & & & \\ 
 IRA Account &  & 4.294e-04$^{***}$ & 4.160e-06 &  &  \\ 
  &  & (5.948e-05) & (4.650e-06) &  &  \\ 
  & & & & & \\ 
 Keogh Account &  & 1.567e-03$^{***}$ & 2.390e-05 &  &  \\ 
  &  & (2.464e-04) & (1.924e-05) &  &  \\ 
  & & & & & \\ 
 Margin Account &  & $-$3.729e-03$^{***}$ & $-$3.400e-06 &  &  \\ 
  &  & (8.506e-05) & (6.660e-06) &  &  \\ 
  & & & & & \\ 
 Schwab Account &  & 3.173e-03$^{***}$ & 2.750e-06 &  &  \\ 
  &  & (6.037e-05) & (4.750e-06) &  &  \\ 
  & & & & & \\ 
 General Brokerage &  & $-$8.599e-03$^{***}$ & $-$1.880e-05$^{***}$ &  &  \\ 
  &  & (5.810e-05) & (4.580e-06) &  &  \\ 
  & & & & & \\ 
 Active Trader &  & $-$4.582e-03$^{***}$ & $-$4.604e-05$^{***}$ &  &  \\ 
  &  & (8.123e-05) & (6.400e-06) &  &  \\ 
  & & & & & \\ 
 Num. of Stocks &  &  & 1.378e-05$^{***}$ &  & 7.771e-05$^{***}$ \\ 
  &  &  & (5.700e-07) &  & (2.580e-06) \\ 
  & & & & & \\ 
 Portfolio Sharpe Ratio &  &  & 1.000e-08 &  & 2.700e-07 \\ 
  &  &  & (3.000e-07) &  & (5.700e-07) \\ 
  & & & & & \\ 
 Portfolio Expected Return &  &  & 0.523$^{***}$ &  & 0.45$^{***}$ \\ 
  &  &  & (1.086e-04) &  & (3.010e-04) \\ 
  & & & & & \\ 
 Portfolio Std. Deviation  &  &  & $-$0.838$^{***}$ &  & $-$0.82$^{***}$ \\ 
  &  &  & (5.242e-05) &  & (2.002e-04) \\ 
  & & & & & \\ 
 VIX Index &  &  &  & $-$3.581e-05$^{***}$ & $-$3.863e-04$^{***}$ \\ 
  &  &  &  & (4.380e-06) & (1.290e-06) \\ 
  & & & & & \\ 
\hline \\[-1.8ex] 
Observations & 1841177 & 1841177 & 1841177 & 1841177 & 1841177 \\ 
R$^{2}$ & 0.024 & 0.041 & 0.994 & 3.739e-05 & 0.915 \\ 
Adjusted R$^{2}$ & 0.024 & 0.041 & 0.994 & $-$0.029 & 0.912 \\ 
F Statistic & 1.698e+03$^{***}$  & 2.409e+03$^{***}$  & 8.463e+06$^{***}$  & 66.884$^{***}$  & 3.840e+06$^{***}$ \\ 
& (df = 27; 1841079) & (df = 33; 1841073) & (df = 37; 1841069) & (df = 1; 1788885) &  (df = 5; 1788881) \\
\hline 
\hline \\[-1.8ex] 
\textit{Note:}  & \multicolumn{5}{r}{$^{*}$p$<$0.1; $^{**}$p$<$0.05; $^{***}$p$<$0.01} \\ 
\end{longtable}  
\end{landscape}
\normalsize
\pagebreak

\begin{landscape}
\pagestyle{empty}
\setlength\LTleft{0.2in}
\setlength\LTright{-1in plus 1 fill}
\tiny
\begin{longtable}{@{\extracolsep{\fill}}lcccc} 
  \caption{Panel Regression on Efficiency with Time Fixed Effects \label{table:eff_twoway}} \\
\\[-1.8ex]\hline 
\hline \\[-1.8ex] 
 & \multicolumn{4}{c}{\textit{Dependent variable:}} \\ 
\cline{2-5} 
\\[-1.8ex] & \multicolumn{4}{c}{Portfolio Efficiency} \\ 
 & (1) & (2) & (3) & (4)  \\ 
\hline \\[-1.8ex] 
 Num. of Stocks & 3.589e-03$^{***}$ & 3.589e-03$^{***}$ & 3.510e-03$^{***}$ & 2.919e-05$^{***}$ \\ 
  & (7.840e-06)  & (7.840e-06) & (7.730e-06) & (1.320e-06) \\ 
  & & & & \\ 
 Portfolio Sharpe Ratio &   & 1.829e-05$^{***}$ & 8.290e-06$^{***}$ & $-$1.000e-08 \\ 
  &   & (1.820e-06) & (1.800e-06) & (2.900e-07) \\ 
  & & & & \\ 
 Portfolio Expected Return &   &  & 0.221$^{***}$ & 0.514$^{***}$ \\ 
  &   &  & (9.429e-04) & (1.561e-04) \\ 
  & & & & \\ 
 Portfolio Std. Deviation &   &  &  & $-$0.835$^{***}$ \\ 
  &   &  &  & (1.019e-04) \\ 
  & & & & \\ 
\hline \\[-1.8ex] 
Observations & 1841177  & 1841177 & 1841177 & 1841177 \\ 
R$^{2}$ & 0.105  & 0.105 & 0.132 & 0.977 \\ 
Adjusted R$^{2}$ & 0.079  & 0.079 & 0.106 & 0.977 \\ 
F Statistic & 2.097e+05$^{***}$ (df = 1; 1788815)  & 1.049e+05$^{***}$ (df = 2; 1788814) & 9.048e+04$^{***}$ (df = 3; 1788813) & 1.939e+07$^{***}$ (df = 4; 1788812) \\ 
\hline 
\hline \\[-1.8ex] 
\textit{Note:}  & \multicolumn{4}{r}{$^{*}$p$<$0.1; $^{**}$p$<$0.05; $^{***}$p$<$0.01} \\ 
\end{longtable}  
\end{landscape}
\normalsize
\pagebreak

\appendix
\section{Proof of Proposition \ref{prop}}
The portfolios on the efficient frontier are weighted linear combinations of the $p$ risky assets with weights $w = [w_1, \dots, w_p]^T$, which minimize risk for a given level of return.
We denote the return level by $a$, and use $e$ to denote a column vector whose entries are all equal to one. 
The weights on the efficient frontier are recovered as the solution to the following optimization problem:
\begin{equation*}
\begin{aligned}
& \min_{w} w^T \Sigma w\\
\mbox{s.t. } & w^T \mu = a \mbox{ and } w^T e = 1.
\end{aligned}
\end{equation*}

By introducing the vector of Lagrange multipliers $(\lambda, \nu)$, the Lagrangian $L$ can be expressed as
\begin{equation*}
L =  w^T \Sigma w + \lambda (w^T \mu - a) + \nu (w^T e - 1).
\end{equation*}
Using the first order condition, we obtain that the solution to the above optimization problem is equivalent to the solution of the  system of equations:
\begin{equation*}
\begin{aligned}
 2 \Sigma w + \lambda \mu + \nu e & = 0,\qquad \qquad
 w^T\mu &  = a, \qquad \qquad
 w^T e &  = 1.\\
\end{aligned}
\end{equation*}
Such a system can be easily solved, and its solution is given by
\begin{equation*}
\begin{aligned}
 \lambda & = 2 \frac{-aC + B}{AC - B^2},\\
 \nu & = 2 \frac{-A + aB}{AC - B^2}, \\
 w & = \frac{1}{AC - B^2} \Sigma^{-1} ((C \mu - B e) a + (Ae - B \mu)),\\
\end{aligned}
\end{equation*}
where we have introduced $A,B,C$ for notational convenience: 
\begin{equation*}
\begin{aligned}
 A = & \mu^T \Sigma^{-1} \mu,\qquad \qquad
 B = & \mu^T \Sigma^{-1} e = e^T \Sigma^{-1} \mu,\qquad \qquad
 C = & e^T \Sigma^{-1} e.
\end{aligned}
\end{equation*}
Therefore, the portfolios $(\sigma_p, \mu_p)$ on the efficient frontier satisfies 
\begin{equation*}\label{eqn:eff}
\sigma_p^2 = w^T \Sigma w = \frac{1}{AC - B^2} (C \mu_p^2 - 2B \mu_p + A).
\end{equation*}

Next, we use the fact that the capital market line is tangent to the efficient frontier, and the market portfolio is on the efficient frontier as well as on the capital market line. We first take the derivative of $\sigma_p$ with respect to $\mu_p$, and obtain
\begin{equation*}
\begin{aligned}
\frac{\partial \sigma_p}{\partial \mu_p}= & \frac{1}{2} \left(\frac{1}{AC - B^2} (C \mu_p^2 - 2 B \mu_p + A)\right)^{-1/2} \left(\frac{1}{AC - B^2} ( 2 C \mu_p - 2B) \right)\\
= & \frac{C \mu_{p} - B}{(AC - B^2)^{1/2} (C \mu_{p}^2 - 2B \mu_{p} + A)^{1/2}}.
\end{aligned}
\end{equation*}
Then, we evaluate this derivative at $(\mu_{mkt},\sigma_{mkt})$, and set it equal to the inverse slope of the capital market line, i.e.,
\begin{equation*}
\begin{aligned}
& \frac{C \mu_{mkt} - B}{(AC - B^2)^{1/2} (C \mu_{mkt}^2 - 2B \mu_{mkt} + A)^{1/2}} = \frac{\left(\frac{1}{AC - B^2} (C \mu_{mkt}^2 - 2B \mu_{mkt} + A) \right)^{1/2}}{\mu_{mkt - r_f}}.\\
\end{aligned}
\end{equation*}
After solving the quadratic equation, we obtain $\mu_{mkt}$, and $\sigma_{mkt}$ from \eqref{eqn:eff}, i.e.,
\begin{equation*}
	\begin{aligned}
		\mu_{mkt} = & \frac{A - B r_f}{B - C r_f},\qquad \qquad
		\sigma_{mkt} = &  \frac{\sqrt{A - 2B r_f + C r_f^2}}{B -  C r_f}.
	\end{aligned}
\end{equation*}
Using the above expressions for the expected return and volatility of the market portfolio, we can immediately obtain the capital market line, which simply connects the risk-free rate with the market portfolio.

\pagebreak
\bibliographystyle{jf}
\bibliography{master}
\addcontentsline{toc}{section}{Bibliography}

\end{document}